\documentclass[sigconf]{acmart}
\AtBeginDocument{%
  }

\usepackage{enumerate}
\usepackage[utf8]{inputenc}
\usepackage{tablefootnote}
\usepackage{algorithm2e}
\usepackage{graphicx}
\usepackage{xcolor}
\usepackage{multirow}
\usepackage{framed}
\usepackage{url}
\usepackage{xurl}
\usepackage{mathtools}
\usepackage{hyperref}
\usepackage{tikz}
\usepackage{amsmath}
\usepackage{pifont}
\usepackage{algpseudocode}
\usepackage[utf8]{inputenc}
\usepackage{amsthm}
\usepackage{textcomp,booktabs}
\usepackage{tabularx}

\usepackage[caption=false,font=normalsize,labelfont=sf,textfont=sf]{subfig}

\usepackage{flushend}

\newcommand{\newmaterial}[1]{\textcolor{black}{{#1}}}

\copyrightyear{2025}
\acmYear{2025}
\setcopyright{acmlicensed}
\acmConference[HT 2025]{Proceedings of the 36th ACM Conference on Hypertext and Social Media}{September 15--18, 2025}{Chicago, IL, USA}
\acmBooktitle{Proceedings of the 36th ACM Conference on Hypertext and Social Media (HT 2025), September 15--18, 2025, Chicago, IL, USA}
\acmDOI{10.1145/3720553.3746680}
\acmISBN{979-8-4007-1534-1/2025/09}

\begin{document}
\title{Toxicity in State Sponsored Information Operations}

\author{Ashfaq Ali Shafin}
\orcid{0000-0002-0135-0091} 
\affiliation{
  \institution{Florida International University}
  \city{Miami}
  \state{FL}
  \country{USA}
}
\email{shafinashfaqali21@gmail.com}

\author{Khandaker Mamun Ahmed}
\orcid{0000-0002-4713-188X} 
\affiliation{
  \institution{Dakota State University}
  \city{Madison}
  \state{SD}
  \country{USA}
}
\email{khandakermamun.ahmed@dsu.edu}

\begin{abstract}
State-sponsored information operations (IOs) increasingly influence global discourse on social media platforms, yet their emotional and rhetorical strategies remain inadequately characterized in scientific literature. This study presents the first comprehensive analysis of toxic language deployment within such campaigns, examining 56 million posts from over 42 thousand accounts linked to 18 distinct geopolitical entities on X/Twitter. Using Google's Perspective API, we systematically detect and quantify six categories of toxic content and analyze their distribution across national origins, linguistic structures, and engagement metrics, providing essential information regarding the underlying patterns of such operations. Our findings reveal that while toxic content constitutes only 1.53\% of all posts, they are associated with disproportionately high engagement and appear to be strategically deployed in specific geopolitical contexts. Notably, toxic content originating from Russian influence operations receives significantly higher user engagement compared to influence operations from any other country in our dataset. Our code is available at {\color{magenta}\url{https://github.com/shafin191/Toxic_IO}}.
\end{abstract}



\maketitle
\vspace{-7pt}
\section{Introduction}

State-controlled information operations, also known as influence operations (IOs), have emerged as powerful tools for shaping public discourse, building support, and advancing geopolitical goals through social media platforms. These operations often employ coordinated amplification, inauthentic behavior, and manipulative narratives to influence online conversations and public perception~\cite{LPBF24, RCHS23, LGF20, VET20, MLFF25}. While prior research has predominantly examined influence operations through network analysis, behavioral patterns, and automation detection, the rhetorical and emotional dimensions of these operations remain significantly understudied.


One particularly potent yet underexplored rhetorical strategy is the deployment of toxic language including hate speech, threats, profanity, insults, and identity-based attacks by state-linked IO operators (also known as operatives~\cite{RCHS23}, IO drivers~\cite{LPBF24}, or trolls~\cite{LGF20})\footnote{In this paper, we use the terms \textit{operators} and \textit{operatives} interchangeably to refer to accounts/users who participated in information/influence operations.
}. Toxic discourse can deepen social polarization, provoke emotional reactions, and justify harmful actions. For instance, during the Rohingya refugee crisis and the Russia–Ukraine conflict, toxic narratives were widely circulated to rationalize violence and mobilize public sentiment~\cite{S20,GBPF23}. Despite such evidence, we lack a systematic understanding of how toxic rhetoric functions within state-backed influence operations, whether it is intentionally deployed or emerges as a byproduct of broader manipulation tactics.

In this study, we present the first large-scale, cross-national analysis of toxic language in state-linked information operations. We analyze over 56 million posts from 42,405 operators attributed to 18 geopolitical entities, using X/Twitter’s publicly released archive of state-sponsored operations. Leveraging Google’s Perspective API~\cite{perspective_api}, we detect six categories of toxic content and examine their distribution across countries, post structure, and user engagement. Our study is guided by two research questions:
\vspace{-1pt}
\begin{itemize}
    \item \textbf{RQ1:} How frequently do influence operatives post toxic content, and how does this vary across countries?
    \item \textbf{RQ2:} What structural and engagement-related differences exist between toxic and non-toxic posts?
\end{itemize}
\vspace{-1pt}

Toxic posts represent only a small portion (1.53\%) of the dataset. However, approximately one-third of influence operators posted at least one toxic post. Importantly, toxic posts receive more engagement than non-toxic ones. Toxicity appears to be used deliberately by operators originating from countries such as Russia, Venezuela, and Cuba to enhance visibility and provoke reactions. Our findings suggest that toxicity is not uniformly applied across state-led campaigns, but rather functions as a selective rhetorical device. This work contributes to the broader understanding of how toxicity is used in influence campaigns and underscores the need for content moderation approaches that account for rhetorical strategy.

\section{Related Work}
\noindent{{\bf Information Operations (IOs)}}. Information operations on social media involve coordinated attempts to manipulate public discourse, often through tactics such as disinformation, amplification, and the use of deceptive or fake accounts. Existing literature has focused on distinguishing influence operators from organic users~\cite{LPBF24, MLFF25, VET20, LGF20}, often through behavioral, temporal, or network-based features. Other works have explored campaign tactics such as coordinated retweeting, content manipulation, and cross-platform influence~\cite{PFM20, VNCCGL24}. Despite this growing body of research, most work has prioritized structural detection and classification, rather than exploring the rhetorical strategies embedded in these operations. In particular, the use of toxic language, a key mechanism for emotional manipulation and provocation, remains underexamined.

\noindent{{\bf Toxic Content on Social Media}}.
Toxic language in online spaces has been widely studied for its effects on user behavior, engagement, and community norms. Prior works have shown that toxic content spreads more rapidly and reaches larger audiences than non-toxic content, especially when posted by influential users~\cite{MDGM19,MPF24}. For instance, hateful posts on fringe platforms like Gab were found to propagate more quickly than non-hateful ones~\cite{MDGM19}, and toxic posts from verified accounts on X/Twitter exhibited greater persistence and virality~\cite{MPF24}. A previous study examined the conditions under which users produce hate speech, highlighting the role of anonymity and platform design in facilitating harmful expression~\cite{MSB17}. While these studies offer extensive insights into toxic discourse in general contexts, relatively few studies investigate its use within organized, state-linked operations. Our work contributes to this area by analyzing how toxicity is deployed by influence operators across geopolitical contexts, and by quantifying its structural and engagement-related properties.

\noindent{{\bf Engagement in Toxic Content}}.
\newmaterial{Information operations strategically exploit well-established psychological mechanisms to maximize engagement and reach. Negativity bias, an evolutionary adaptation for threat detection, causes humans to allocate disproportionate attention to negative information compared to positive content~\cite{BBFV01}. This cognitive bias is amplified by emotional arousal, where high-activation negative emotions such as anger and anxiety increase sharing likelihood compared to low-arousal emotions like sadness~\cite{BM12}. Influence operations weaponize these psychological vulnerabilities by disseminating toxic content~\cite{SA25}, and the toxicity serves as an engagement amplifier by increasing user interaction and platform activity, and exploiting algorithmic systems that prioritize high-engagement content~\cite{BJMS22}. This dynamic provides malicious actors with a critical asymmetry: the ability to achieve disproportionate reach using minimal resources.}

\section{Data}
We use the publicly released state-sponsored information operations dataset published by X/Twitter, which documents coordinated state-controlled influence campaigns conducted on the platform that violated its policies on platform manipulation and coordinated inauthentic behavior between 2018 and 2021~\cite{IOTwitter}. The dataset offers a comprehensive view of these operations, providing both the complete timeline (all tweets and retweets) and account metadata for identified influence operators. Our analysis focuses on content attributed to actors from 18 distinct geopolitical entities, representing a broad global distribution. These include major state actors such as China, Iran, Russia, Venezuela, as well as politically contentious regions such as Catalonia. In total, the dataset includes 42,405 unique influence operators, which collectively posted over 56 million distinct posts, including both original tweets and retweets. 
\begin{table}[t!]
    \centering
    \resizebox{\columnwidth}{!}{
        \begin{tabular}{lrrrr}
            \toprule
            \textbf{Country} & \textbf{\# Influence} & \textbf{\# IO} & \textbf{\# Toxic} & \textbf{\#Toxic} \\
            \textbf{} & \textbf{Operators} & \textbf{Posts} & \textbf{Operators(\%)} & \textbf{Posts(\%)} \\
            \midrule
            Armenia (AM) & 31 & 68,481 & 7(22.58\%) & 174(0.25\%) \\
            Bangladesh (BD) & 8 & 2,824 & 4(50.00\%) & 28(0.99\%) \\
            Catalonia (CT) & 59 & 1,168 & 17(28.81\%) & 33(2.82\%) \\
            China (CN) & 7,301 & 5,250,954 & 1,334(18.27\%) & 95,508(1.81\%) \\
            Cuba (CU) & 502 & 4,564,491 & 414(82.47\%) & 58,476(1.28\%) \\
            Honduras (HN) & 2,994 & 1,049,578 & 1,034(34.54\%) & 27,612(2.63\%) \\
            Indonesia (ID) & 641 & 706,705 & 162(25.27\%) & 6,178(0.87\%) \\
            Iran (IR) & 4,383 & 5,214,603 & 2,216(50.58\%) & 124,331(2.38\%) \\
            Mexico (MX) & 264 & 17,970 & 186(70.45\%) & 942(5.24\%) \\
            Russia (RU) & 1,470 & 3,342,694 & 1,179(80.27\%) & 176,645(5.28\%) \\
            Saudi Arabia (SA) & 4,820 & 21,926,672 & 2,001(41.53\%) & 76,063(0.35\%) \\
            Serbia (RS) & 7,478 & 1,768,131 & 2,096(28.03\%) & 17,550(0.99\%) \\
            Spain (ES) & 216 & 54,421 & 132(61.11\%) & 1,099(2.02\%) \\
            Tanzania (TZ) & 134 & 2,544 & 18(13.43\%) & 67(2.63\%) \\
            Turkey (TR) & 4,251 & 475,261 & 1,465(34.47\%) & 11,121(2.34\%) \\
            UAE (AE) & 3,606 & 1,170,452 & 1,606(44.53\%) & 9,734(0.83\%) \\
            Uganda (UG) & 388 & 440,940 & 256(65.98\%) & 15,508(3.52\%) \\
            Venezuela (VE) & 1,859 & 10,275,358 & 1,425(76.63\%) & 238,218(2.32\%) \\
            \midrule
            \textbf{Total} & \textbf{42,405} & \textbf{56,359,247} & \textbf{14,200(33.47\%)} & \textbf{859,285(1.53\%)} \\
            \bottomrule
        \end{tabular}
    }
    \caption{Overview of IO posts and operators from 18 regions, highlighting toxic posts and operators. On average, Russian operators spread the most toxic content (5.28\%).}
    \vspace{-25pt}
    \label{tab:IOdata}
\end{table}


Table~\ref{tab:IOdata} summarizes the dataset by region, including the number of influence operators and IO posts. Serbia had the most operators ($n = 7{,}478$), while Saudi Arabia generated the highest number of posts ($n = 21{,}926{,}672$). On average, each country had 2{,}244.72 operators ($SD = 2{,}515.57$) and 3{,}129{,}624.83 posts ($SD = 5{,}441{,}444.25$).

\section{Methodology}

\begin{figure}[h]
    \centering
    \includegraphics[width=\linewidth]{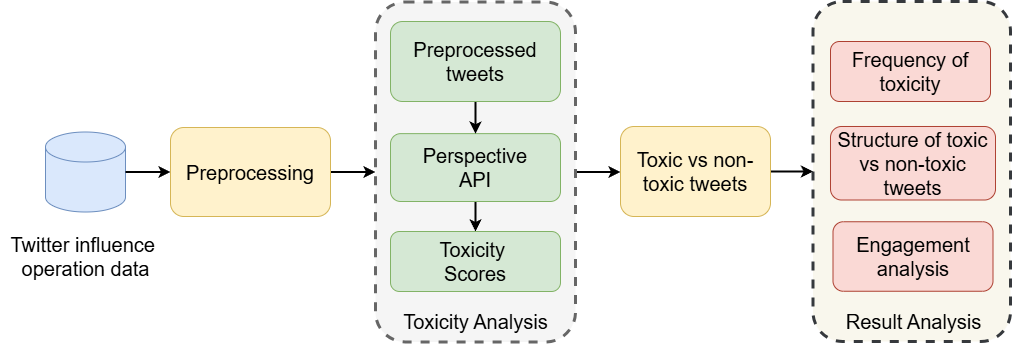}
    \vspace{-15pt}
    \caption{Architecture for Toxicity Analysis of State-Sponsored Information Operations on Twitter.}
    \label{fig:Architecture}
    \vspace{-0 pt}
\end{figure}

The overall methodology for toxicity analysis of influence operations is illustrated in Figure~\ref{fig:Architecture}. First, we preprocessed all post texts in our dataset to ensure clean and consistent input for automated toxicity analysis. This cleaning process involved removing hyperlinks, user mentions (e.g., \texttt{@username}), HTML codes, emojis, and extraneous white spaces. We also normalized the text by converting it to lowercase and standardizing common contractions and repeated characters to improve the accuracy of toxicity scoring.

\vspace{-5pt}

\vspace{-7pt}

After preprocessing, we employed Google’s Perspective API~\cite{perspective_api} to evaluate the toxicity of each post. The API returned probability scores (ranging from 0 to 1) for six different attributes: \textit{toxicity}, \textit{severe toxicity}, \textit{identity attack}, \textit{insult}, \textit{profanity}, and \textit{threat}. A post was marked as \textit{toxic} for a given attribute if its score exceeded 0.5 in any of the six categories, otherwise it was marked as \textit{non-toxic}. Since these categories are not mutually exclusive, a single post could be associated with multiple attributes. 

\begin{figure*}[h!]
    \centering
    \includegraphics[width=0.91\linewidth]{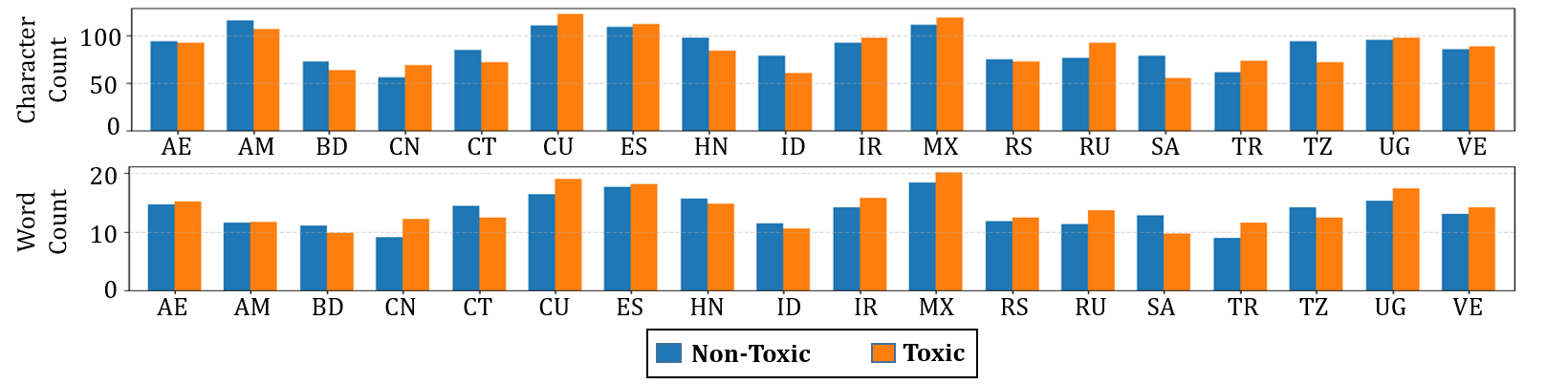}
    \vspace{-12pt}
    \caption{Comparison of post length between toxic and non-toxic posts across countries. In most countries, toxic posts tend to contain more characters and words than their non-toxic counterparts. }
    \label{fig:tweet_length}
    \vspace{-11pt}
\end{figure*}

To ensure the reliability of the automatic annotations, we randomly selected a balanced subset of 500 English posts (50\% toxic and 50\% non-toxic) for manual verification by two independent annotators with experience in social media content moderation. Each annotator reviewed the posts and labeled them as toxic or non-toxic based on the same definitions used by the Perspective API. We calculated inter-rater agreement using Cohen’s Kappa coefficient and observed a high level of consistency between the two raters (Cohen’s $\kappa = 0.81$, agreement = 90.4\%). \newmaterial{Agreement between annotator-1 and the Perspective API was 89.7\% (Cohen’s $\kappa = 0.79$), while agreement between annotator-2 and the Perspective API was 92.8\% (Cohen’s $\kappa = 0.85$).} This manual verification enhanced confidence in the validity of the automated labeling approach and facilitated refinement of edge cases in subsequent analyses.

\section{Findings}

\begin{figure*}[h!]
    \centering
    \includegraphics[width=0.92\linewidth]{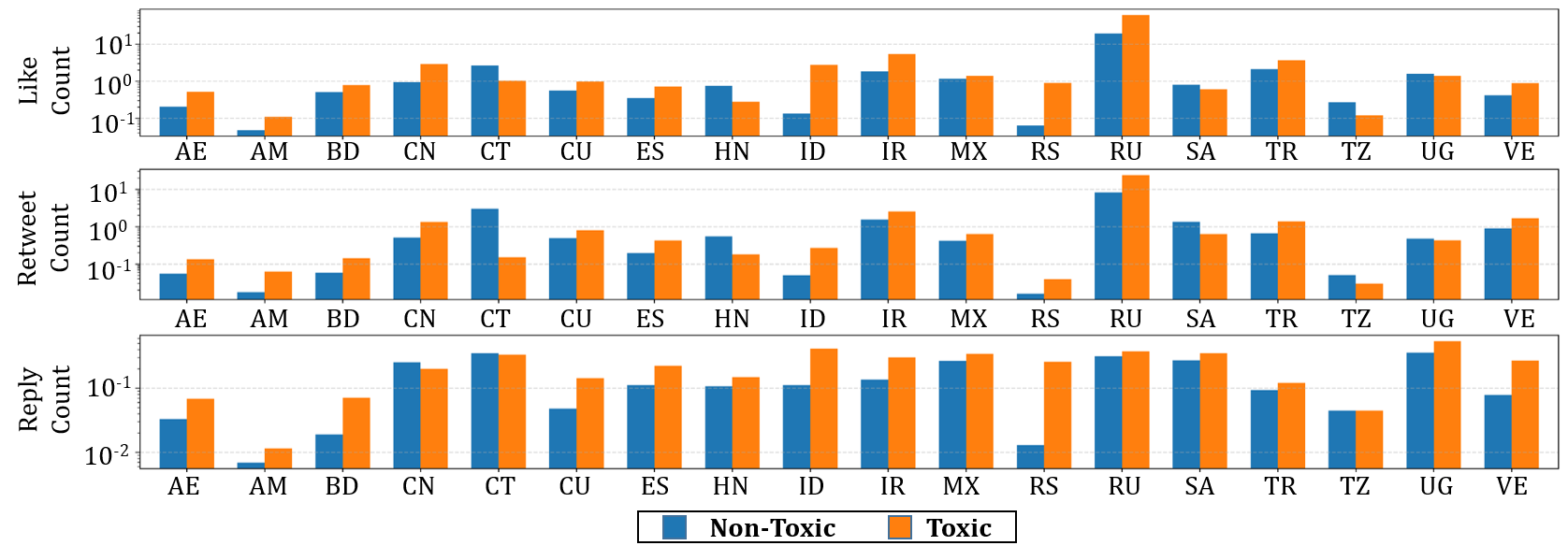}
    \vspace{-12pt}
    \caption{Comparison of post engagement between toxic and non-toxic posts across countries. On average, toxic posts received higher engagement than non-toxic posts in the majority of regions.}
    \vspace{-10pt}
    \label{fig:tweet_engagement}
\end{figure*}

\noindent{\bf{RQ1: Frequency of Toxic Content in IOs}}.
Table~\ref{tab:IOdata} presents a comprehensive summary of the influence operation (IO) dataset analyzed in this study. The dataset analysis identified a substantial presence of toxic content across multiple categories. Out of a total of 859,285 unique posts classified as toxic, the distribution across specific toxicity categories was as follows: 573,763 posts (66.8\%) contained general toxicity, 61,110 posts (7.1\%) exhibited severe toxicity, 344,208 posts (40.1\%) contained profanity, 113,022 posts (13.2\%) included identity attacks, 435,307 posts (50.7\%) contained insults, and 90,495 posts (10.5\%) contained threats. On average, influence operators disseminate toxic content relatively infrequently, with only 1.53\% (859,285 out of 56.3 million) of all posts classified as toxic. Nevertheless, approximately one-third of influence operators (14,200 out of 42,405) have shared toxic content at least once. At the post-level, Russia (5.28\%), Mexico (5.24\%), and Uganda (3.52\%) demonstrate the highest proportions of toxic content, indicating a greater reliance on inflammatory or harmful rhetoric in their operations. Conversely, countries such as Armenia (0.25\%), Saudi Arabia (0.35\%), and Indonesia (0.87\%) display significantly lower rates of toxic posts, suggesting a preference for less aggressive communication strategies. From a user-level perspective, Cuba, Russia, and Venezuela reveal the highest concentration of toxic operators, with 82.47\%, 80.27\%, and 76.63\% of their respective operators posting at least one toxic post. This pattern contrasts markedly with countries like Tanzania (13.43\%) and China (18.27\%), where toxic behavior remains confined to a smaller proportion of operators, despite potentially high overall post volumes.

Among the countries with more than one million posts, we observe substantial variation in toxic content strategies. Russia, Venezuela, and Cuba are notable for generating both a high volume of content and elevated levels of toxicity. Russian operatives generated over 3.3 million posts, with 5.28\% classified as toxic and more than 80\% of operatives exhibiting toxic behavior. Similarly, Cuba and Venezuela each generated millions of posts (4.5M and 10.2M, respectively), with more than 75\% of their operatives producing toxic content. In contrast, China (5.2M posts) and Saudi Arabia (21.9M posts), despite conducting some of the largest operations, demonstrated lower toxicity rates (1.81\% and 0.35\%) and smaller proportions of toxic operatives (18.27\% and 41.53\%, respectively). 

\newmaterial{The variation in toxic content across countries reflects distinct strategic contexts and operational constraints. Russia's high toxicity rates align with documented aggressive disinformation campaigns targeting Western democracies, where polarization serves clear strategic goals such as undermining electoral integrity, amplifying racial tensions, and eroding trust in democratic institutions~\cite{DSRSMFAJ18, LW20}. Conversely, China and Saudi Arabia show lower toxicity despite massive scale, reflecting different strategic priorities focused on supportive narrative promotion rather than disruption~\cite{SA25}. 
This suggests that while toxic content is not universally prevalent in IOs, its use is a key tactic for certain countries aiming to stir controversy, polarize audiences, or achieve political objectives.}

\noindent{\bf{RQ2: Post Structure and Engagement}}. Our quantitative analysis reveals a statistically significant difference in textual dimensions between toxic and non-toxic posts. Toxic content demonstrates greater linguistic density, with a mean character count of 86.98 and average word count of 14.02, compared to non-toxic content which exhibits means of 81.90 characters and 12.84 words, respectively. These findings suggest that toxic digital discourse is characterized by relatively higher verbosity across linguistic parameters. In examining the geographic distribution of these linguistic patterns in Figure~\ref{fig:tweet_length}, we observed consistent trends across national boundaries. For two-thirds of the studied regions, 
toxic posts consistently contained a higher word count than their non-toxic counterparts. Interestingly, character length distributions demonstrated greater homogeneity across different national contexts. The differential between toxic and non-toxic content in terms of character count remained minimal in most countries, indicating that while toxic content tends to employ more words, this does not necessarily correlate with proportional increases in overall character utilization.

We found substantial disparities in engagement metrics between toxic and non-toxic posts within state-sponsored information operations, as illustrated in Figure~\ref{fig:tweet_engagement}. On average, toxic posts received significantly higher engagement across all measures: approximately 14.30 likes, 6.04 reposts, and 0.29 replies per post, totaling 20.63 total engagements on average. In contrast, non-toxic posts garnered only 1.86 likes, 1.42 reposts, and 0.19 replies, totaling 3.46 engagements on average. A two-sample t-test conducted at the post level confirms that this difference is statistically significant (t = 94.11, p < 0.001), indicating that toxic posts consistently attract more engagement than their non-toxic counterparts. These findings raise critical concerns about the disproportionate visibility and potential amplification of harmful content in influence campaigns.

Further, we find toxic operatives (those who posted at least one toxic post) receive significantly more engagement on average than non-toxic operatives. Specifically, toxic operatives averaged 1.70 likes, 0.81 retweets, and 0.22 replies per post, amounting to a total average engagement of 2.72. In contrast, non-toxic operatives received only 0.95 likes, 0.47 retweets, and 0.25 replies per post, with a total average engagement of 1.67. A two-sample t-test confirms that this difference is statistically significant (t = 2.62, p < 0.01), suggesting that operatives involved in toxic behavior tend to attract greater audience interaction overall. This result highlights the potential incentives and visibility benefits that may drive toxic participation in information operations.

We observe a notable divergence in engagement between toxic and non-toxic posts across countries, highlighting potential strategic uses of toxicity in information operations. In countries such as Russia, Iran, China, and Turkey, toxic posts significantly outperformed non-toxic ones in average engagement, with Russia showing the starkest contrast: an average of 86.08 engagements for toxic content versus 27.88 for non-toxic. Similarly, in Indonesia, toxic posts (Mean, $M = 3.45$) garnered more than ten times the engagement of non-toxic posts ($M = 0.29$). These disparities suggest that in certain geopolitical contexts, toxic content may be more likely to provoke interaction or amplification. However, the trend was not universal. For instance, in regions like Honduras, Saudi Arabia, and Catalonia, non-toxic posts had higher engagement, indicating that toxicity is not a uniformly effective strategy.

\vspace{-4pt}
\section{Limitations}
We used Google's Perspective API to identify toxic content, which introduces several methodological constraints. The API supports only a limited subset of languages, consequently restricting our analysis to posts composed in these supported languages. This limitation necessitated the exclusion of a substantial corpus of posts from our influence operation datasets, particularly those written in linguistically diverse contexts such as Persian and Bangla, which currently fall outside the API's analytical capabilities. Research has demonstrated that the API exhibits systematic bias across different languages, with studies showing it misreads certain languages as more toxic than others, particularly non-English languages like German~\cite{NPCLTG25}. This language-specific bias may disproportionately affect toxicity measurements for non-English posts, potentially skewing cross-country comparisons where different languages dominate the discourse. Our manual annotation of 500 English posts helps validate the results, but further multilingual validation is needed.

\section{Conclusion}
This study presents the first comprehensive analysis of toxic language in state-sponsored information operations across 18 countries. While toxic content constitutes only a small fraction of the dataset (42,405 operatives, 56 million posts), its strategic significance is evident with one-third of operatives posting toxic content at least once. Notably, toxic posts generate substantially higher engagement than non-toxic content, suggesting toxicity appears to play a strategic role in amplification mechanisms. Geopolitical distribution reveals marked variation, with Russia demonstrating high proportions of toxic content, while China and Saudi Arabia maintain high volume but relatively low toxicity rates. Toxic posts exhibit greater linguistic density across national contexts, with engagement disparities creating clear incentives for inflammatory rhetoric. These findings indicate toxicity serves as a tactical element within broader influence strategies, necessitating detection systems and platform governance approaches that incorporate behavioral patterns and geopolitical context beyond content-level analysis.

\begin{acks}

We thank the \textbf{Perspective API} team for the extended quota provision that enabled this comprehensive analysis. The authors acknowledge using generative AI tools for grammatical and stylistic improvements only; all research contributions are original. All conceptual and analytical contributions are solely those of the authors. 

\end{acks}

\bibliographystyle{ACM-Reference-Format}
\bibliography{hatespeech, IO}


\begin{thebibliography}{21}


\ifx \showCODEN    \undefined \def \showCODEN     #1{\unskip}     \fi
\ifx \showISBNx    \undefined \def \showISBNx     #1{\unskip}     \fi
\ifx \showISBNxiii \undefined \def \showISBNxiii  #1{\unskip}     \fi
\ifx \showISSN     \undefined \def \showISSN      #1{\unskip}     \fi
\ifx \showLCCN     \undefined \def \showLCCN      #1{\unskip}     \fi
\ifx \shownote     \undefined \def \shownote      #1{#1}          \fi
\ifx \showarticletitle \undefined \def \showarticletitle #1{#1}   \fi
\ifx \showURL      \undefined \def \showURL       {\relax}        \fi
\providecommand\bibfield[2]{#2}
\providecommand\bibinfo[2]{#2}
\providecommand\natexlab[1]{#1}
\providecommand\showeprint[2][]{arXiv:#2}

\bibitem[per(2017)]%
        {perspective_api}
 \bibinfo{year}{2017}\natexlab{}.
\newblock \bibinfo{title}{{Perspective API}}.
\newblock \bibinfo{howpublished}{\url{https://www.perspectiveapi.com}}.
\newblock
\newblock
\shownote{Accessed: 2025-01-06}.


\bibitem[Baumeister et~al\mbox{.}(2001)]%
        {BBFV01}
\bibfield{author}{\bibinfo{person}{Roy~F Baumeister}, \bibinfo{person}{Ellen Bratslavsky}, \bibinfo{person}{Catrin Finkenauer}, {and} \bibinfo{person}{Kathleen~D Vohs}.} \bibinfo{year}{2001}\natexlab{}.
\newblock \showarticletitle{Bad is stronger than good}.
\newblock \bibinfo{journal}{\emph{Review of General Psychology}} \bibinfo{volume}{5}, \bibinfo{number}{4} (\bibinfo{year}{2001}), \bibinfo{pages}{323--370}.
\newblock
\href{https://doi.org/10.1037/1089-2680.5.4.323}{doi:\nolinkurl{10.1037/1089-2680.5.4.323}}


\bibitem[Beknazar-Yuzbashev et~al\mbox{.}(2022)]%
        {BJMS22}
\bibfield{author}{\bibinfo{person}{George Beknazar-Yuzbashev}, \bibinfo{person}{Rafael Jim{\'e}nez-Dur{\'a}n}, \bibinfo{person}{Jesse McCrosky}, {and} \bibinfo{person}{Mateusz Stalinski}.} \bibinfo{year}{2022}\natexlab{}.
\newblock \showarticletitle{Toxic Content and User Engagement on Social Media: Evidence from a Field Experiment}.
\newblock \bibinfo{journal}{\emph{Available at SSRN}} (\bibinfo{date}{November} \bibinfo{year}{2022}).
\newblock
\href{https://doi.org/10.2139/ssrn.4307346}{doi:\nolinkurl{10.2139/ssrn.4307346}}


\bibitem[Berger and Milkman(2012)]%
        {BM12}
\bibfield{author}{\bibinfo{person}{Jonah Berger} {and} \bibinfo{person}{Katherine~L. Milkman}.} \bibinfo{year}{2012}\natexlab{}.
\newblock \showarticletitle{What Makes Online Content Viral?}
\newblock \bibinfo{journal}{\emph{Journal of Marketing Research}} \bibinfo{volume}{49}, \bibinfo{number}{2} (\bibinfo{year}{2012}), \bibinfo{pages}{192--205}.
\newblock
\href{https://doi.org/10.1509/jmr.10.0353}{doi:\nolinkurl{10.1509/jmr.10.0353}}


\bibitem[DiResta et~al\mbox{.}(2018)]%
        {DSRSMFAJ18}
\bibfield{author}{\bibinfo{person}{Renee DiResta}, \bibinfo{person}{Kris Shaffer}, \bibinfo{person}{Becky Ruppel}, \bibinfo{person}{David Sullivan}, \bibinfo{person}{Robert Matney}, \bibinfo{person}{Ryan Fox}, \bibinfo{person}{Jonathan Albright}, {and} \bibinfo{person}{Ben Johnson}.} \bibinfo{year}{2018}\natexlab{}.
\newblock \bibinfo{booktitle}{\emph{The Tactics \& Tropes of the Internet Research Agency}}.
\newblock \bibinfo{type}{{T}echnical {R}eport}.
\newblock
\urldef\tempurl%
\url{https://digitalcommons.unl.edu/senatedocs/2/}
\showURL{%
\tempurl}


\bibitem[Geissler et~al\mbox{.}(2023)]%
        {GBPF23}
\bibfield{author}{\bibinfo{person}{Dominique Geissler}, \bibinfo{person}{Dominik B{\"a}r}, \bibinfo{person}{Nicolas Pr{\"o}llochs}, {and} \bibinfo{person}{Stefan Feuerriegel}.} \bibinfo{year}{2023}\natexlab{}.
\newblock \showarticletitle{Russian propaganda on social media during the 2022 invasion of Ukraine}.
\newblock \bibinfo{journal}{\emph{EPJ Data Science}} \bibinfo{volume}{12}, \bibinfo{number}{1} (\bibinfo{year}{2023}), \bibinfo{pages}{35}.
\newblock


\bibitem[Linvill and Warren(2020)]%
        {LW20}
\bibfield{author}{\bibinfo{person}{Darren~L. Linvill} {and} \bibinfo{person}{Patrick~L. Warren}.} \bibinfo{year}{2020}\natexlab{}.
\newblock \showarticletitle{Troll Factories: Manufacturing Specialized Disinformation on Twitter}.
\newblock \bibinfo{journal}{\emph{Political Communication}} \bibinfo{volume}{37}, \bibinfo{number}{4} (\bibinfo{year}{2020}), \bibinfo{pages}{447--467}.
\newblock
\href{https://doi.org/10.1080/10584609.2020.1718257}{doi:\nolinkurl{10.1080/10584609.2020.1718257}}
\showeprint{https://doi.org/10.1080/10584609.2020.1718257}


\bibitem[Luceri et~al\mbox{.}(2020)]%
        {LGF20}
\bibfield{author}{\bibinfo{person}{Luca Luceri}, \bibinfo{person}{Silvia Giordano}, {and} \bibinfo{person}{Emilio Ferrara}.} \bibinfo{year}{2020}\natexlab{}.
\newblock \showarticletitle{Detecting Troll Behavior via Inverse Reinforcement Learning: A Case Study of Russian Trolls in the 2016 US Election}.
\newblock \bibinfo{journal}{\emph{Proceedings of the International AAAI Conference on Web and Social Media}} \bibinfo{volume}{14}, \bibinfo{number}{1} (\bibinfo{date}{May} \bibinfo{year}{2020}), \bibinfo{pages}{417--427}.
\newblock
\href{https://doi.org/10.1609/icwsm.v14i1.7311}{doi:\nolinkurl{10.1609/icwsm.v14i1.7311}}


\bibitem[Luceri et~al\mbox{.}(2024)]%
        {LPBF24}
\bibfield{author}{\bibinfo{person}{Luca Luceri}, \bibinfo{person}{Valeria Pant\`{e}}, \bibinfo{person}{Keith Burghardt}, {and} \bibinfo{person}{Emilio Ferrara}.} \bibinfo{year}{2024}\natexlab{}.
\newblock \showarticletitle{Unmasking the Web of Deceit: Uncovering Coordinated Activity to Expose Information Operations on Twitter}. In \bibinfo{booktitle}{\emph{Proceedings of the ACM Web Conference 2024}} (Singapore, Singapore) \emph{(\bibinfo{series}{WWW '24})}. \bibinfo{publisher}{Association for Computing Machinery}, \bibinfo{address}{New York, NY, USA}, \bibinfo{pages}{2530–2541}.
\newblock
\showISBNx{9798400701719}
\href{https://doi.org/10.1145/3589334.3645529}{doi:\nolinkurl{10.1145/3589334.3645529}}


\bibitem[Maarouf et~al\mbox{.}(2024)]%
        {MPF24}
\bibfield{author}{\bibinfo{person}{Abdurahman Maarouf}, \bibinfo{person}{Nicolas Pr\"{o}llochs}, {and} \bibinfo{person}{Stefan Feuerriegel}.} \bibinfo{year}{2024}\natexlab{}.
\newblock \showarticletitle{The Virality of Hate Speech on Social Media}.
\newblock \bibinfo{journal}{\emph{Proc. ACM Hum.-Comput. Interact.}} \bibinfo{volume}{8}, \bibinfo{number}{CSCW1}, Article \bibinfo{articleno}{186} (\bibinfo{date}{April} \bibinfo{year}{2024}), \bibinfo{numpages}{22}~pages.
\newblock
\href{https://doi.org/10.1145/3641025}{doi:\nolinkurl{10.1145/3641025}}


\bibitem[Mathew et~al\mbox{.}(2019)]%
        {MDGM19}
\bibfield{author}{\bibinfo{person}{Binny Mathew}, \bibinfo{person}{Ritam Dutt}, \bibinfo{person}{Pawan Goyal}, {and} \bibinfo{person}{Animesh Mukherjee}.} \bibinfo{year}{2019}\natexlab{}.
\newblock \showarticletitle{Spread of Hate Speech in Online Social Media}. In \bibinfo{booktitle}{\emph{Proceedings of the 10th ACM Conference on Web Science}} (Boston, Massachusetts, USA) \emph{(\bibinfo{series}{WebSci '19})}. \bibinfo{publisher}{Association for Computing Machinery}, \bibinfo{address}{New York, NY, USA}, \bibinfo{pages}{173–182}.
\newblock
\showISBNx{9781450362023}
\href{https://doi.org/10.1145/3292522.3326034}{doi:\nolinkurl{10.1145/3292522.3326034}}


\bibitem[Minici et~al\mbox{.}(2025)]%
        {MLFF25}
\bibfield{author}{\bibinfo{person}{Marco Minici}, \bibinfo{person}{Luca Luceri}, \bibinfo{person}{Francesco Fabbri}, {and} \bibinfo{person}{Emilio Ferrara}.} \bibinfo{year}{2025}\natexlab{}.
\newblock \showarticletitle{IOHunter: Graph Foundation Model to Uncover Online Information Operations}.
\newblock \bibinfo{journal}{\emph{Proceedings of the AAAI Conference on Artificial Intelligence}} \bibinfo{volume}{39}, \bibinfo{number}{27} (\bibinfo{date}{Apr.} \bibinfo{year}{2025}), \bibinfo{pages}{28258--28266}.
\newblock
\href{https://doi.org/10.1609/aaai.v39i27.35046}{doi:\nolinkurl{10.1609/aaai.v39i27.35046}}


\bibitem[Mondal et~al\mbox{.}(2017)]%
        {MSB17}
\bibfield{author}{\bibinfo{person}{Mainack Mondal}, \bibinfo{person}{Leandro~Ara\'{u}jo Silva}, {and} \bibinfo{person}{Fabr\'{\i}cio Benevenuto}.} \bibinfo{year}{2017}\natexlab{}.
\newblock \showarticletitle{A Measurement Study of Hate Speech in Social Media}. In \bibinfo{booktitle}{\emph{Proceedings of the 28th ACM Conference on Hypertext and Social Media}} (Prague, Czech Republic) \emph{(\bibinfo{series}{HT '17})}. \bibinfo{publisher}{Association for Computing Machinery}, \bibinfo{address}{New York, NY, USA}, \bibinfo{pages}{85–94}.
\newblock
\showISBNx{9781450347082}
\href{https://doi.org/10.1145/3078714.3078723}{doi:\nolinkurl{10.1145/3078714.3078723}}


\bibitem[Nogara et~al\mbox{.}(2025)]%
        {NPCLTG25}
\bibfield{author}{\bibinfo{person}{Gianluca Nogara}, \bibinfo{person}{Francesco Pierri}, \bibinfo{person}{Stefano Cresci}, \bibinfo{person}{Luca Luceri}, \bibinfo{person}{Petter Törnberg}, {and} \bibinfo{person}{Silvia Giordano}.} \bibinfo{year}{2025}\natexlab{}.
\newblock \showarticletitle{Toxic Bias: Perspective API Misreads German as More Toxic}.
\newblock \bibinfo{journal}{\emph{Proceedings of the International AAAI Conference on Web and Social Media}} \bibinfo{volume}{19}, \bibinfo{number}{1} (\bibinfo{date}{Jun.} \bibinfo{year}{2025}), \bibinfo{pages}{1346--1357}.
\newblock
\href{https://doi.org/10.1609/icwsm.v19i1.35876}{doi:\nolinkurl{10.1609/icwsm.v19i1.35876}}


\bibitem[Pacheco et~al\mbox{.}(2020)]%
        {PFM20}
\bibfield{author}{\bibinfo{person}{Diogo Pacheco}, \bibinfo{person}{Alessandro Flammini}, {and} \bibinfo{person}{Filippo Menczer}.} \bibinfo{year}{2020}\natexlab{}.
\newblock \showarticletitle{Unveiling Coordinated Groups Behind White Helmets Disinformation}. In \bibinfo{booktitle}{\emph{Companion Proceedings of the Web Conference 2020}} (Taipei, Taiwan) \emph{(\bibinfo{series}{WWW '20})}. \bibinfo{publisher}{Association for Computing Machinery}, \bibinfo{address}{New York, NY, USA}, \bibinfo{pages}{611–616}.
\newblock
\showISBNx{9781450370240}
\href{https://doi.org/10.1145/3366424.3385775}{doi:\nolinkurl{10.1145/3366424.3385775}}


\bibitem[Recabarren et~al\mbox{.}(2023)]%
        {RCHS23}
\bibfield{author}{\bibinfo{person}{Ruben Recabarren}, \bibinfo{person}{Bogdan Carbunar}, \bibinfo{person}{Nestor Hernandez}, {and} \bibinfo{person}{Ashfaq~Ali Shafin}.} \bibinfo{year}{2023}\natexlab{}.
\newblock \showarticletitle{Strategies and vulnerabilities of participants in venezuelan influence operations}. In \bibinfo{booktitle}{\emph{32nd USENIX Security Symposium (USENIX Security 23)}}. \bibinfo{pages}{6683--6700}.
\newblock


\bibitem[Roth(2019)]%
        {IOTwitter}
\bibfield{author}{\bibinfo{person}{Yoel Roth}.} \bibinfo{year}{2019}\natexlab{}.
\newblock \bibinfo{title}{Information operations on Twitter: principles, process, and disclosure}.
\newblock \bibinfo{howpublished}{\url{https://blog.x.com/en_us/topics/company/2019/information-ops-on-twitter }}.
\newblock
\newblock
\shownote{Accessed: April - 29 - 2025}.


\bibitem[Shafin and Ahmed(2025)]%
        {SA25}
\bibfield{author}{\bibinfo{person}{Ashfaq~Ali Shafin} {and} \bibinfo{person}{Khandaker~Mamun Ahmed}.} \bibinfo{year}{2025}\natexlab{}.
\newblock \bibinfo{title}{The Language of Influence: Sentiment, Emotion, and Hate Speech in State Sponsored Influence Operations}.
\newblock
\showeprint[arxiv]{2505.07212}~[cs.SI]
\urldef\tempurl%
\url{https://arxiv.org/abs/2505.07212}
\showURL{%
\tempurl}


\bibitem[Siddiquee(2020)]%
        {S20}
\bibfield{author}{\bibinfo{person}{Md.~Ali Siddiquee}.} \bibinfo{year}{2020}\natexlab{}.
\newblock \showarticletitle{The portrayal of the Rohingya genocide and refugee crisis in the age of post-truth politics}.
\newblock \bibinfo{journal}{\emph{Asian Journal of Comparative Politics}} \bibinfo{volume}{5}, \bibinfo{number}{2} (\bibinfo{year}{2020}), \bibinfo{pages}{89--103}.
\newblock
\href{https://doi.org/10.1177/2057891119864454}{doi:\nolinkurl{10.1177/2057891119864454}}
\showeprint{https://doi.org/10.1177/2057891119864454}


\bibitem[Vargas et~al\mbox{.}(2020)]%
        {VET20}
\bibfield{author}{\bibinfo{person}{Luis Vargas}, \bibinfo{person}{Patrick Emami}, {and} \bibinfo{person}{Patrick Traynor}.} \bibinfo{year}{2020}\natexlab{}.
\newblock \showarticletitle{On the Detection of Disinformation Campaign Activity with Network Analysis}. In \bibinfo{booktitle}{\emph{Proceedings of the 2020 ACM SIGSAC Conference on Cloud Computing Security Workshop}} (Virtual Event, USA) \emph{(\bibinfo{series}{CCSW'20})}. \bibinfo{publisher}{Association for Computing Machinery}, \bibinfo{address}{New York, NY, USA}, \bibinfo{pages}{133–146}.
\newblock
\showISBNx{9781450380843}
\href{https://doi.org/10.1145/3411495.3421363}{doi:\nolinkurl{10.1145/3411495.3421363}}


\bibitem[Vishnuprasad et~al\mbox{.}(2024)]%
        {VNCCGL24}
\bibfield{author}{\bibinfo{person}{Padinjaredath~Suresh Vishnuprasad}, \bibinfo{person}{Gianluca Nogara}, \bibinfo{person}{Felipe Cardoso}, \bibinfo{person}{Stefano Cresci}, \bibinfo{person}{Silvia Giordano}, {and} \bibinfo{person}{Luca Luceri}.} \bibinfo{year}{2024}\natexlab{}.
\newblock \showarticletitle{Tracking Fringe and Coordinated Activity on Twitter Leading Up to the US Capitol Attack}.
\newblock \bibinfo{journal}{\emph{Proceedings of the International AAAI Conference on Web and Social Media}} \bibinfo{volume}{18}, \bibinfo{number}{1} (\bibinfo{date}{May} \bibinfo{year}{2024}), \bibinfo{pages}{1557--1570}.
\newblock
\href{https://doi.org/10.1609/icwsm.v18i1.31409}{doi:\nolinkurl{10.1609/icwsm.v18i1.31409}}


\end{thebibliography}

\end{document}